\documentclass[aps,prd,preprintnumbers,showpacs]{revtex4}
\usepackage[dvips]{graphicx}
\begin{document}

%
%

\preprint{Nisho-1-2012}
\title{Quark Pair Production in Expanding Glasma}
\author{Aiichi Iwazaki}
\address{International Economics and Politics, Nishogakusha University,\\ 
6-16 3-bantyo Tiyoda Tokyo 102-8336, Japan.}   
\date{September 4, 2011. Revised: January 13 2012}
\begin{abstract}
Glasma in high energy heavy ion collisions
is longitudinal classical color electric and magnetic fields. 
The color electric field has been shown to produces quark and anti-quark pairs
by the Schwinger mechanism
and to oscillate with time in non-expanding glasma, that is, plasma oscillation.
On the other hand,  in the expanding glasma
we show that the field decreases with the plasma oscillation.
We can explicitly obtain the solutions representing such temporal
behaviors in the system with $\tau$ and $\eta$ coordinates.
We show these results by using massless QED as a simplified model of QCD.   
\end{abstract}
\hspace*{0.3cm}
\hspace*{1cm}

\maketitle

\section{introduction}
Initial states of color gauge fields (glasma) produced immediately after high energy heavy-ion collisions
have recently received much attention. 
The gauge fields are longitudinal classical color electric $E$ and magnetic $B$ fields.
The presence of such classical gauge fields 
has been discussed on the basis of a fairly reliable
effective theory of QCD at high energy, that is, a model of color glass condensate (CGC)\cite{gl,cgc}.
The glasma is generated by small x gluons in nuclei.
It is expected that
the decay of the glasma leads to thermalized quark gluon plasma (QGP). 
The study of the decay of the glasma is very significant
because the classical and quantum behaviors of the coherent non-Abelian gauge fields have not 
been experimentally known. 
 
The glasma is homogeneous in the longitudinal
direction but inhomogeneous in the transverse directions. Hence,
we may view that it forms electric and magnetic flux tubes extending in the longitudinal direction. 
Their field strengths are given by the saturation momentum $Q_s$ in high energy heavy ion collisions
such that $eB\sim Q_s^2$ and $eE \sim Q_s^2$; $e(>0)$  denotes the coupling constant.
In the previous papers\cite{iwa,itakura,hii} we have shown a possibility that 
the famous Nielsen-Olesen instability\cite{nielsen} makes the color magnetic field $B$
decay.  The possibility has been partially confirmed by
the comparison between our results and numerical simulations\cite{venugopalan,berges}. 
Such decay is a first step toward the generation of QGP.
( The gauge fields produced initially in the model of color glass condensate have very random
spatial distribution without no typical length scales\cite{fukushima}. The Nielsen-Olesen unstable modes
do not arise under such random configulations of the fields.
But soon after the production they would rapidly evolve
to form a smooth distribution with the typical length scale $Q_s^{-1}$ owing to the nonlinear interactions
among the fields\cite{berg}. Then, the unstable modes can arise under the smooth background gauge fields. )

Furthermore,
we have also discussed\cite{iwazaki1,iwazaki2,old} the decay of the color electric field; the decay
is caused by Schwinger mechanism\cite{schwinger}, that is, the pair production of quarks and anti-quarks.
The mechanism has been extensively explored\cite{tanji} since the discovery of the Klein paradox. 
A new feature in the glasma is that it is composed of not only electric field
but also magnetic field collinear to the electric field. Under such a circumstance
there are only several studies of the mechanism involving the back reaction of 
the produced particles on the electric field\cite{iwazaki2,tanji,suga}.  
It has been found that the number density of
quarks and the electric field oscillate in time, namely the plasma oscillation
when the quarks are free after their production. On the other hand, it has recently shown\cite{iwazaki3} that
when the quarks interact with each other in heat bath,
the electric field decays without the oscillation, while their number density increases and is
saturated. These results have been obtained in a system of non-expanding glasma
using QED for simplicity. 

In this paper we analyze the behaviors of the electric field and the quark number density in the expanding glasma.
We only consider the case of non-interacting quarks after their production.
The pair production in the expanding glasma has already been discussed\cite{lap}, but
there is still no analysis in the case that both electric and magnetic fields are present. 
In order to describe the expanding glasma, we choose
a coordinate system such as $\tau=\sqrt{t^2-z^2}$ and $\eta=\log\sqrt{(t+z)/(t-z)}$
instead of $t$ and $z$ in Cartesian coordinate. 
( $\tau$ is a proper time associated with a particle whose longitudinal coordinate is specified by
$\eta=$ constant. ) 
For simplicity, we analyze U(1) gauge theory instead of SU(3) gauge theory.
Thus, we use the terminology of electrons and positrons.
They are assumed to be massless particles. This is because the    
corresponding masses of quarks can be neglected in the glasma. The typical scales
in the glasma is given by
the saturation momentum $Q_s$, which
is much larger than the mass of the quarks ; $Q_s\simeq 1\rm{GeV} \sim 2\rm{GeV}$ for RICH or LHC. 
We also assume that only relevant states in the pair production are those in the lowest Landau level
with their energies given by $|p_z|$ in the Cartesian coordinate where $p_z$ is the momentum parallel to $\vec{B}$.
The assumption holds only in the limit of $eB \to \infty $, although both $eB$ and $eE$ are of the order of 
$Q_s^2$ in the real glasma. We discuss that the assumption is approximately correct 
even when $eB$ and $eE$ are the same order of magnitudes as each other. 
We find that both the electron number density 
and the electric field decrease oscillating with time.
The decrease is caused by the expansion of glasma.

In order to show these results, we use the chiral anomaly.
The chiral anomaly is a quite powerful tool for the discussion of the Schwinger mechanism.
It is effective only when both collinear electric and magnetic fields are present.
When the magnetic field is very strong,
the produced particles occupy only the states in the lowest Landau level.
It implies that the transverse motion of the particles are frozen
and only the longitudinal motion is allowed.
Thus, the particle production effectively takes place in two dimensional space-time.
Then, the form of the chiral anomaly is much simplified. 
The pair production can be discussed without detail calculations
but simply by solving the anomaly equation.

In the next section \ref{22} we explain how the pair production 
effectively arises in two dimensional space-time when the 
magnetic field is very strong.
We define a two dimensional field operator using creation and annihilation
operators of the states in the lowest Landau level. Electric charge and
chiral current of the states are represented with the field operator.
The analysis in this section is performed in the Cartesian coordinate. 
In the section \ref{3}
we show the utility of the chiral anomaly for the pair production 
by using the field operator.
In the section \ref{4} we apply our method to analyse the pair production in 
$\tau$ and $\eta$ coordinates. We find that the electric field decreases oscillating 
with time owing to the back reaction and the expansion of the glasma.
We summarize our results in the section \ref{5}.


\vspace{0.2cm} 

\section{the lowest Landau level and reduction of spatial dimensions}
\label{22}
 
First, we explain solutions of Dirac equation under the homogeneous magnetic field in the Cartesian coordinate
and the reduction of the four dimensional operators  to the two dimensional ones. 
We use the gauge potential $\vec{A}=(0,xB,0)$ of the magnetic field $\vec{B}=(0,0,B)$.
Then, the solutions representing massless fermions in the lowest Landau level
which we are only concerned with, 
are given in the following,

\begin{equation}
\label{sol}
\Psi=N_0\exp(-iE_pt+ip_y y+ip_z z)\exp\Bigl(-\frac{eB}{2}(x-\frac{p_y}{eB})^2\Bigr)u(p) \quad
\mbox{with} \quad u(p)=\left(
\begin{array}{@{\,}cccc@{\,}}
1\\0\\\ p_z/E_p\\0 \end{array}\right)\quad
\end{equation}
where $E_p=\pm |p_z|$ represents the energy and
$N_0$ denotes a normalization constant.
On the other hand, the states in higher Landau levels have the energies $\pm\sqrt{enB+p_z^2}$
where the integer $n \ge 1$ denotes Landau level. Thus, in the limit of $eB \to \infty $,
only relevant states in the pair production are the states with the energies $|p_z|$ 
in the lowest Landau level.
( In addition to the states there are other states in the lowest Landau level
whose energies are given by $\sqrt{eB+p_z^2}$.  The states are those with
magnetic moments anti-parallel to $\vec{B}$, carrying much higher energies than
the energies of the above states. These states are also irrelevant in the limit. )
Hereafter we only consider the states in the lowest Landau level, which
are the most relevant states in the Schwinger mechanism when $B\gg E$.

The transverse motions are described by the factor $\exp(+ip_y y)\exp(-eB(x-p_y/eB)^2/2)$
of the wave functions in the lowest Landau level. 
The states are degenerate in the momentum $p_y$.
The states in higher Landau levels are described by different functions of the transverse
coordinates $x$ and $y$. In the sense, 
the transverse motions are frozen, 
as long as we are concerned with such states in the lowest Landau level.
Only dynamical behaviors of the physical quantities are allowed in the longitudinal direction.
Namely,
the states in the lowest Landau level are specified with quantum numbers $p_z$ and $p_y$,
whose energies are given by $|p_z|$. Since the electric field parallel to $\vec{B}$
accelerates the particles, only the momentum $p_z$ increases or decreases, but the momentum $p_y$
does not change. Thus, the mometum $p_y$
is conserved and 
the particles do not jump into higher Landau levels as long as 
their energies $|p_z| \le eB$. ( Although the momentum $p_z$ increases by the acceleration of the electric field,
the electric field decays with the back reaction until $p_z$ becomes sufficiently large such as $|p_z|>eB$. )  
Namely, the transverse motions specified by the coordinates $x$ and $y$ are frozen.
Therefore, we only need to specify the quantum number $p_z$ in order to
describe the states in the lowest Landau level.

Since they are trivial in the transverse directions,
we take an average over the transverse space of physical quantities, e.g. $\bar{\Psi} O \Psi$ with
gamma matrices $O$. Then, we find that the average over the transverse space of $\bar{\Psi} O \Psi$
are given by $\bar{\psi} O \psi$  
where the field operator $\psi$ is defined as,

\begin{equation}
\label{2}
\psi=\int \frac{dp}{4\pi}\Bigr(\exp(-i\omega_p t+ipz)u(p)A_p+\exp(i\omega_p-ipz)u(p)B_p^{\dagger}\Bigr)
\end{equation}  
with $\omega_p\equiv |p|$ and $p\equiv p_z$,
where $A_p$ and $B_p$ denote annihilation operators of particles and anti-particles
with the momentum $p$, respectively,
and satisfy the following commutation relations,

\begin{equation}
A_pA_q^{\dagger}+A_q^{\dagger}A_p\equiv\{A_p,A_q^{\dagger}\}=\{B_p,B_q^{\dagger}\}=2\pi \delta(p-q),
 \quad \mbox{ the other commutators vanish. }
\end{equation}
Then, it follows that

\begin{equation}
\{\psi^{\dagger} (t,z),\psi(t,z')\}=\delta(z-z').
\end{equation}
The field $\psi$ satisfies effectively 2 dimensional Dirac equation, 
$(i\gamma^0\partial_t+\gamma^3\partial_z)\psi=0$ where 

\[ \gamma^0 = \left(
\begin{array}{@{\,}cccc@{\,}}
1& 0\\0&-1 \end{array}\right), \quad  \gamma^3 = \left(
\begin{array}{@{\,}cccc@{\,}}
0& \sigma_3\\-\sigma_3 &0 \end{array}\right),
\quad \mbox{and} \quad
\gamma^5 = \left(
\begin{array}{@{\,}cccc@{\,}}
0& 1\\ 1 &0 \end{array}\right). \]

Note that the spinor and the gamma matrices are not two dimensional one.
The term of "two dimensional" means that the field depends only on the coordinates $t$ and $z$.
When homogeneous electric field $E=-\partial_t A_3(t)$ is present in addition to the magnetic field,
the operators $A_p$ and $B_p$ depend on $t$, but keep the above commutation relations.
They satisfy the equations,
$i\partial_tA_p=eA_3A_p \,p/|p|$ and $i\partial_tB_p^{\dagger}=eA_3 B_p^{\dagger} \,p/|p|$.

Using the field $\psi$, we find that the expectation values of the electric charge $e\bar{\psi} \gamma^0 \psi $, 
electric current $e\bar{\psi} \gamma^3 \psi $, chiral charge
$\bar{\psi} \gamma^0 \gamma^5 \psi $ and 
chiral current $\bar{\psi} \gamma^3 \gamma^5 \psi $ are given respectively by

\begin{eqnarray}
\label{J}
J^0&\equiv & \langle e\bar{\psi} \gamma^0 \psi \rangle=e\int \frac{dp}{2\pi}(n_p-\tilde{n}_p), \quad \mbox{and} \quad
J^3\equiv \langle e\bar{\psi} \gamma^3 \psi \rangle=e\int \frac{dp}{2\pi}\frac{p}{\omega_p} (n_p-\tilde{n}_p) \\ \nonumber
J^5 &\equiv &\langle \bar{\psi} \gamma^0 \gamma^5 \psi \rangle=\int \frac{dp}{2\pi}\frac{p}{\omega_p}(n_p-\tilde{n}_p)
\quad \mbox{and} \quad 
J^{3,5}\equiv \langle \bar{\psi} \gamma^3\gamma^5 \psi \rangle=\int \frac{dp}{2\pi}(n_p-\tilde{n}_p),
\end{eqnarray}
where operator products are defined as normal ordered product with respect to $A_p$ and $B_p$. 
We have taken the expectation values of these quantities by assuming 
appropriate states $|\,\,\rangle $ relevant to the particle production
under the electric field, which satisfy

\begin{equation}
\langle A_p^{\dagger}A_q \rangle =2\pi \delta(p-q)n_p(t), \quad 
\langle B_p^{\dagger}B_q \rangle =2\pi \delta(p-q)\tilde{n}_p(t)  \quad \mbox{and} \quad
\langle A_p^{\dagger}B_q^{\dagger} \rangle=\langle B_p A_q \rangle=0
\end{equation}

Obviously $A_p$ ( $B_p$ ) represents 
annihilation operators of positrons ( electrons ) with
the electric charge $ e > 0$ ( $-e < 0$ ); $n_p(t)$ represents momentum distribution of the positrons,
while $\tilde{n}_p(t)$ represents that of electrons.
Both of them depend on time $t$, in general. 

We denote the number density of positrons 
by $N_p=\int \frac{dp}{2\pi}n_p$ and of electrons by
$N_e=\int \frac{dp}{2\pi}\tilde{n}_p$ respectively.
Then, we find that 

\begin{eqnarray}
\label{7}
J^0&=&eN_p-eN_e,\quad J^{3,5}=N_p-N_e,\nonumber \\ 
J^3&=&eJ^5=e(N_p+N_e) \quad \mbox{for} \quad \vec{E}\cdot\vec{B}>0
\quad \mbox{and} \quad J^3=eJ^5=-e(N_p+N_e) \quad \mbox{for}
\quad \vec{E}\cdot\vec{B}<0.
\end{eqnarray}

The electric charge $J^0$ is given by the number densities of
electrons and positrons as expected. On the other hand we need to explain
why the form of the chirality $J^5$ is given as in Eq(\ref{7}). 
Because positrons ( electrons ) are accelerated into the direction
parallel ( anti-parallel ) to $\vec{E}$, the momentum of positrons ( electrons ) 
is given by $p>0$ ( $p<0$ ) for $\vec{E}\cdot\vec{B}>0$ or $p<0$ ( $p>0$ )
for $\vec{E}\cdot\vec{B}<0$; $\vec{B}=(0,0,B)$ with $B>0$.
( The particles are spontaneously produced with their momentum equal to $0$
because of the Pauli principle. ) Then, 
the helicity of both positrons and electrons is positive for 
$\vec{E}\cdot\vec{B}>0$ or negative for $\vec{E}\cdot\vec{B}<0$,
since the spins of the positrons ( electrons )
are pointed into the direction parallel ( anti-parallel ) to $\vec{B}$. 
Therefore, the chirality $J^5$ is given as in Eq(\ref{7}).
This is owing to the fact that the produced particles occupy the lowest Landau level with their 
magnetic moment parallel to $\vec{B}$ and their momenta are determined by the electric field.


\vspace{0.2cm}
\section{chiral anomaly and plasma oscillation}
\label{3}

Next, we explain the utility of the chiral anomaly and how 
the pair production under homogeneous electric and magnetic fields is described by using the anomaly. 
Suppose that the magnetic field is sufficiently strong for the charged particles to occupy only the lowest Landau level. Then,
we may use the effectively two dimensional foumulas as described above. 
The anomaly equation is given in the following,

\begin{equation}
\label{chiral}
 \partial_t J^5=\frac{e^2}{2\pi^2}\vec{E}(t)\cdot\vec{B}
\end{equation}
where we have taken an average over the transverse space $\vec{x}_T=(x_1,x_2)$,
of the chiral current so that $\int d\vec{x}_T \partial_{\vec{T}}J^{\vec{T},5}=0$.
We have also used the relation $\partial_3 J^{3,5}=0$.
The term in the right hand side of Eq(\ref{chiral}) arises by taking account of quantum effects ( loop diagrams ) of 
electrons. That is, although the chiral current is conserved classically,
the conservation is violated by the quantum effects. It is important to note that
the violation term is given by the product of the electric and magnetic fields.

When the electric field $E$ is switched on at $t=0$ in vacuum, the pair production arises 
and the chiral charge $J^5$ is produced
according to the anomaly equation (\ref{chiral}). Suppose that the electric field is
parallel to $B$. Then, positrons move to the direction parallel to $E$, while
electrons move to the direction anti-parallel to $E$. We have $n_p \propto \theta (p)$ for positrons 
and $\tilde{n}_p\propto \theta(-p)$  
for electrons. Thus, it follows from Eq(\ref{J}) that $J^5=N_p+N_e$. Since $N_p=N_e$ in the pair production,
the anomaly equation becomes 
\begin{equation}
\label{chiral2}
\partial_tN=s E
\end{equation}
with $N\equiv N_p$ and $s\equiv e^2B/4\pi^2$.
Similarly, when $E$ is anti-parallel to $B$, then
$n_p\propto \theta(-p)$ and $\tilde{n}_p\propto \theta(p)$. Thus, $J^5=-N_p-N_e$.
Since $\vec{E}\cdot\vec{B}=-EB$, the anomaly equation (\ref{chiral2}) is also obtained.
Hence it holds in both cases with $\vec{E}\cdot\vec{B}=EB$ and $\vec{E}\cdot\vec{B}=-EB$.

Obviously, the number density of the electrons and positrons produced by the electric field
is governed by the anomaly equation (\ref{chiral2}). For example, when $E$ is switched on at $t=0$,
the particles are produced so that the number density is given by $N=s Et$.
This is the result with no back reaction of the charged particles on $E$.
In order to take into account of the back reaction, we need to solve 
a Maxwell equation $\partial_tE=-J^3=-2eN$ as well as the anomaly equation (\ref{chiral2});
the Maxwell equation holds in a spatially homogeneous system ( $\rm{rot}\,\vec{B}=0$ ). 
From these equations, we derive the equation 

\begin{equation}
\label{n}
\partial_t^2E(t)+2es E(t)=\partial_t^2E(t)+\frac{e^3B}{2\pi^2}E(t)=0,
\end{equation}
whose solutions with the initial conditions $E(t=0)=E_0$ and $N(t=0)=0$ 
are trivially obtained; $E=E_0\cos(\sqrt{2es}\,t)$ and $N=E_0\sqrt{s/2e}\sin(\sqrt{2es}\,t)$. 
The electric field shows a plasma oscillation in Fig.1 with the frequency 
of the order of $\sqrt{eB}$ ( $\sim Q_s$ in the glasma ).
Although we have solved only classical equations, the anomaly equation involves
all of quantum effects associated with the pair productions.
Thus, the chiral anomaly is very useful tool for the investigation of the Schwinger mechanism
when both strong magnetic and electric field are present.

\begin{figure}[htb]
 \centering
 \includegraphics*[width=65mm]{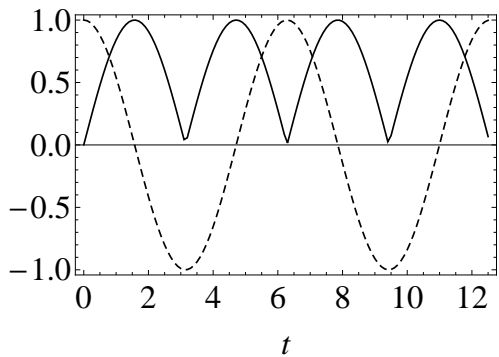}
 \caption{electric field $E(t)$ (dashed line)
and number density of electrons $N(t)$ (solid line) with arbitrary scale}
    \label{l2ea4-f1}
\end{figure}

We should make a comment on the scales involved in the discussion.
The above result ( the plasma oscillation ) holds in the limit of $ eB \gg eE $ where
only the states in the lowest Landau level are relevant. On the other hand when $eB \ll eE $,
the states in higher Landau levels become important 
since the energy $|p_F(t)|=|\int_0^t dt' eE|$ of electrons can be
much larger than $\sqrt{eB}$ soon after their production.
In the glasma we have $eB \simeq eE\sim Q_s^2 \gg m_{\rm{quark}}^2$ ( the square of quark mass ). 
We argue that even in such a case 
the states in higher Landau levels are still not important according to the following reason.
Most of the electrons are produced with their energies equal to $0$. After their production,
they are accelerated and their energies increase.  The energies of the electrons produced at $t=0$ reach the energy 
$|p_F|=\int_0^{Q_s^{-1}} dt' eE \simeq eE Q_s^{-1}\sim \sqrt{eB}$
at the time of $Q_s^{-1}$, the typical time scale of the oscillation. Then,
they may make a transition to a state in a higher Landau level with the energy $\sqrt{eB}$.
Such electrons are those which are produced just after the electric field
is switched on. Most of electrons still have their energies less than $\sqrt{eB}$.
So, the number of the electrons whose energies reach the critical energy $\sqrt{eB}$ is still a fraction. 
Most of the electrons stay in the lowest Landau level.

It has recently been shown\cite{suga} that the vacuum decay by the pair production
is mainly caused by massless particles 
in the lowest Landau level even if $eE\sim eB$. 
Although their analysis does not involve the back reaction of the particles
produced, it supports 
our approximation that the relevant states for 
the decay of the electric field are those in the lowest
Landau level.

\vspace{0.2cm}
\section{pair production in expanding glasma}
\label{4}

Up to now we have discussed the pair production in the system 
where the electric and magnetic fields are infinitely extending in space.
This does not correspond to the case of the real glasma,
which are expanding after the heavy ion collisions.
Thus we now proceed to show how $E$ and $N$
behave in a system where the fields and particles expand. Namely, we calculate their behaviors in 
$\tau$ and $\eta $ coordinates in which
the background field $E_{\eta}$ is uniform in $\eta$.
The behaviors of the fermions in the transverse directions 
specified by the coordinates $x$ and $y$ are
identical to those in Cartesian coordinates. 
They are frozen in the states of the lowest Landau level.

We first rewrite the Dirac equation, $\Bigr(\gamma^0(i\partial_t-eA_0)+\gamma^3(i\partial_z-eA_3)\Bigr)\psi=0$
by using the formulas,

\begin{equation}
\partial_0=\partial_0\tau \,\partial_{\tau}+\partial_0\eta\,\partial_{\eta}=
\cosh\eta \,\partial_{\tau}-\frac{\sinh\eta}{\tau}\,\partial_{\eta}, \quad
\partial_z=\partial_z\tau\,\partial_{\tau}+\partial_z\eta\,\partial_{\eta}=
-\sinh\eta\,\partial_{\tau}+\frac{\cosh\eta}{\tau}\,\partial_{\eta}
\end{equation} 
and

\begin{equation}
A_0=\partial_0\tau \,A_{\tau}+\partial_0\eta \,A_{\eta}=\cosh\eta \,A_{\tau}-\frac{\sinh\eta}{\tau}\,A_{\eta},
\quad 
A_z=\partial_z\tau\,A_{\tau}+\partial_z\eta\,A_{\eta}=-\sinh\eta\,A_{\tau} +\frac{\cosh\eta}{\tau}\,A_{\eta}.
\end{equation}
Then, we obtain

\begin{equation}
\label{D}
\Bigr(\gamma^0(i\partial_{\tau}-eA_{\tau})+\frac{1}{\tau}\gamma^3(i\partial_{\eta}-eA_{\eta})\Bigr)\phi=0
\end{equation}
where we have set $\psi\equiv U\phi/\sqrt{\tau}$ with $U\equiv \cosh\frac{\eta}{2}+\gamma^0\gamma^3 \sinh\frac{\eta}{2}$.

We solve the equation(\ref{D}) with $A_{\tau}=0$ by assuming that $A_{\eta}$ depends only on $\tau$.
The assumption corresponds to the fact that the glasma is homogeneous in $\eta $. 
The solutions are given by

\begin{equation}
\phi_{\pm}=\int \frac{dk_{\eta}}{2\pi}
\frac{A_k^{\pm}}{\sqrt{2}}\exp(ik_{\eta}\eta)v_{\pm}f_k^{\pm} \quad \mbox{with} \quad f_k^{\pm}=
\exp(\mp ik_{\eta}\log\tau\,\mp i\Omega(\tau)) \quad \mbox{and} \quad 
v_{\pm}=\left(
\begin{array}{@{\,}cccc@{\,}}
1\\0\\\pm1\\0 \end{array}\right),
\end{equation}
where $A_k^{\pm}$ are constants and $\Omega$ satisfies $\partial_{\tau}\Omega=eA_{\eta}/\tau$.
We have taken into account of the fact that $\phi$ or $\psi$ represents only the states in the lowest
Landau level. 

Using these solutions, 
we define the field operator corresponding to the field operator in eq(\ref{2}),

\begin{equation}
\hat{\phi}=\int \frac{dk_{\eta}}{2\sqrt{2}\pi}\Biggr(\exp(ik_{\eta}\eta)A_k\Bigl(\theta(k_{\eta})v_+f_k^+
+\theta(-k_{\eta})v_-f_k^-\Big)
+\exp(-ik_{\eta}\eta)B_k^{\dagger}\Bigl(\theta(k_{\eta})v_+f_{-k}^++\theta(-k_{\eta}) v_-f_{-k}^-\Bigr)\Biggr ) 
\end{equation}
where the first terms ( $\propto \exp(ik_{\eta}-i|k_{\eta}|\log\tau)$ )
represent the solutions of the positive frequency 
and the second ones ( $\propto \exp(-ik_{\eta}+i|k_{\eta}|\log\tau)$ ) do the solutions of the negative
frequency. 
$A_k$ and $B_k$ satisfy the commutation relations,

\begin{equation}
\{A_k,A_{k'}^{\dagger}\}=\{B_k,B_{k'}^{\dagger}\}=2\pi \delta(k_{\eta}-k'_{\eta}), \quad \mbox{ the other commutators vanish. }
\end{equation}

Using the commutation relations, we find 

\begin{equation}
\{\hat{\phi}_a(\eta, \tau),\hat{\phi}_b^{\dagger}(\eta',\tau)\}=\delta_{a,b}\delta(\eta-\eta'),
\end{equation}
with $a,b=1,\mbox{or},3$.

\vspace{0.2cm}

Now, we express the chiral anomaly in $\tau$ and $\eta$ coordinates.
It is easy to see that

\begin{equation}
\partial_0J^{5,0}+\partial_3J^{5,3}=\partial_{\tau}J^{5,\tau}+\frac{1}{\tau}J^{5,\tau}+\partial_{\eta}J^{5,\eta}
=2s_{\eta} E_{\eta}
\end{equation} 
with $J^{5,a}=\langle \bar{\psi}\gamma^a\gamma^5\psi\rangle$ ( $a=0, 3$ ) 
and $s_{\eta}\equiv e^2B_{\eta}/4\pi^2$,
where the last term $\partial_{\eta}J^{5,\eta} $ is assumed to vanish corresponding 
to the fact that the glasma is homogeneous in $\eta$.
The current $J^{5,\tau}$ can be represented in terms of $J^{5,0}$ and $J^{5,3}$ in the following,

\begin{equation}
J^{5,\tau}=\partial_0\tau J^{5,0}+\partial_3\tau J^{5,3}=\cosh\eta \,J^{5,0}-\sinh\eta \,J^{5,3}
\end{equation}
Hereafter, we consider only the region of the central rapidity $\eta\simeq 0$.
Then, we have 

\begin{equation}
J^{5,\tau}\simeq J^{5,0}=\langle \frac{1}{\tau}\hat{\phi}^{\dagger}\gamma^5 \hat{\phi}\rangle.
\end{equation}

In addition to the anomaly equation, we exploit a Maxwell equation, $\partial_0F^{0,3}=J^3$,
in order to take into account of the back reaction of the charged particles on $E_{\eta}$.
The equation is represented in $\tau$ and $\eta$ coordinates as

\begin{equation}
\partial_{\tau}(\tau F^{\tau,\eta})=\partial_{\tau}\Bigr(\frac{F_{\tau,\eta}}{-\tau}\Bigr)
=\partial_{\tau}E_{\eta}=-\tau J^{\eta}\simeq -J^3
=-\langle \frac{e\hat{\phi}^{\dagger}\gamma^0\gamma^3\hat{\phi}}{\tau}\rangle,
\end{equation}
where we have taken only the region of the central rapidity.

Therefore, we have the following equations to find the temporal behaviors of the electric field
$E_{\eta}$ and the number density of positrons or electrons,

\begin{equation}
\label{eqs}
\partial_{\tau}J^{5,\tau}+\frac{1}{\tau}J^{5,\tau}=2s_{\eta} E_{\eta} \quad \mbox{and} 
\quad \partial_{\tau}E_{\eta}=-J^3=-\langle \frac{e\hat{\phi}^{\dagger}\gamma^0\gamma^3 \hat{\phi}}{\tau}\rangle,
\end{equation} 
with $J^{5,\tau}=\langle\frac{1}{\tau}\hat{\phi}^{\dagger}\gamma^5 \hat{\phi}\rangle$,
where we have assumed that the electric field is parallel to the magnetic field
$\vec{E}\cdot\vec{B}=E_{\eta}B_{\eta}>0$. When the electric field is anti-parallel to the magnetic field,
the anomaly equation becomes $\partial_{\tau}J^{5,\tau}+\frac{1}{\tau}J^{5,\tau}=-2s_{\eta} E_{\eta}$.

\vspace{0.3cm}
In the above equations
the expectation values of the currents $J^{5,\tau}$ and $J^3$ are taken by assuming appropriate states of positrons
and electrons such that,

\begin{eqnarray}
J^{5,\tau}&=&\langle \frac{1}{\tau}\hat{\phi}^{\dagger}\gamma^5 \hat{\phi}\rangle=
\int \frac{dk_{\eta}}{2\pi\tau}\Bigl((n_k-\tilde{n}_k)\theta(k_{\eta})-(n_k-\tilde{n}_k)\theta(-k_{\eta})\Bigr) \\
J^3 
&=&\langle \frac{e\hat{\phi}^{\dagger}\gamma^0\gamma^3 \hat{\phi}}{\tau} \rangle=
\int \frac{dk_{\eta}}{2\pi\tau}e\Bigl((n_k-\tilde{n}_k)\theta(k_{\eta})-(n_k-\tilde{n}_k)\theta(-k_{\eta})\Bigr) ,
\end{eqnarray}
where the state $|\,\, \rangle$ is supposed to satisfy

\begin{equation}
\langle A_k^{\dagger}A_{k'} \rangle =2\pi \delta(k_{\eta}-k'_{\eta})n_k(t), \quad 
\langle B_k^{\dagger}B_{k'} \rangle =2\pi \delta(k_{\eta}-k'_{\eta})\tilde{n}_k(t)  \quad \mbox{and} \quad
\langle A_k^{\dagger}B_{k'}^{\dagger} \rangle=\langle B_k A_{k'} \rangle=0.
\end{equation}

The quantities $\frac{n_k}{\tau}$,
 and $\frac{\tilde{n}_k}{\tau}$
denote the momentum distributions of positrons and electrons, respectively.
This can be easily understood by noting that the electric charge density is given by
$ J^0\equiv \frac{\langle e\bar{\hat{\phi}} \gamma^0 \hat{\phi} \rangle}{\tau}= \int \frac{dk_{\eta}}{2\pi\tau}e(n_k-\tilde{n}_k)$.
Obviously, $J^0=0$ since electric charge is conserved and the initial state has no electrons and positrons.

We should make a comment that when the electric field is parallel to 
the magnetic field, i.e. $\vec{E}\cdot\vec{B}=E_{\eta}B_{\eta}>0$,
positrons ( electrons ) in the pair production have momentum $k_{\eta}>0$ ( $k_{\eta}<0$ ) 
owing to the acceleration by the electric field.
On the other hand, when $\vec{E}\cdot\vec{B}=-E_{\eta}B_{\eta}<0$, positrons have $k_{\eta}<0$ and electrons do $k_{\eta}>0$. 
That is,

\begin{eqnarray}
&&(n_k-\tilde{n}_k)\theta(k_{\eta})=n_k\theta(k_{\eta})\neq 0 \quad \mbox{and} \quad
(n_k-\tilde{n}_k)\theta(-k_{\eta})=-\tilde{n}_k\theta(-k_{\eta})\neq 0
\quad \mbox{for} \quad \vec{E}\cdot\vec{B}>0, \\, 
&&(n_k-\tilde{n}_k)\theta(k_{\eta})=-\tilde{n}_k\theta(k_{\eta})\neq 0 \quad \mbox{and} \quad 
(n_k-\tilde{n}_k)\theta(-k_{\eta})=n_k\theta(-k_{\eta})\neq 0
\quad \mbox{for}\quad \vec{E}\cdot\vec{B}<0.
\end{eqnarray}

Therefore, we find from these equations that the chirality $J^{5,\tau}$ is given by

\begin{equation}
J^{5,\tau}=N+\tilde{N} \quad \mbox{for}\quad \vec{E}\cdot\vec{B}>0
\quad J^{5,\tau}=-(N+\tilde{N}) \quad \mbox{for}\quad \vec{E}\cdot\vec{B}<0,
\end{equation}
where $N=\frac{\int dk_{\eta}n_k}{2\pi\tau}$ and
$\tilde{N}=\frac{\int dk_{\eta}\tilde{n}_k}{2\pi\tau}$ represent the number densities of
positrons and electrons, respectively.

Therefore, we obtain the following equations governing the pair productions,

\begin{equation}
\label{eqs2}
\partial_{\tau}N+\frac{N}{\tau}=s_{\eta} E_{\eta} \quad \mbox{and} \quad
\partial_{\tau}E_{\eta}=-2eN
\end{equation}
where we have taken into account of the fact that the number of positrons is equal to that of
electrons in the pair production, $N=\tilde{N}$.

We can explicitly solve the equations(\ref{eqs2}) to obtain the temporal behaviors of the electric field $E_{\eta}$ and 
the number density of the charged particles. It is easy to see that
$E_{\eta}$ satisfies

\begin{equation}
\partial_{\tau}^2E_{\eta}+\frac{1}{\tau}\partial_{\tau}E_{\eta}+2es_{\eta} E_{\eta}=0,
\end{equation} 
and $N$ is given by $N=-\partial_{\tau}E_{\eta}/2e$.
We obtain the solutions with the initial conditions, $E_{\eta}(\tau=0)=E_0$ and $N(\tau) \,(\tau =0)=0$,

\begin{equation}
E=E_0J_0(\sqrt{2es_{\eta}}\tau) \quad \mbox{and} \quad N=\frac{E_0\sqrt{2es_{\eta}}}{2e} |J_1(\sqrt{2es_{\eta}}\tau)|
\end{equation}
where $J_{0,1}$ denotes Bessel functions.
We have shown the behaviors of these quantities in Fig.2. We can see that 
they decrease oscillating with time $\tau$, while
they simply oscillate without the decrease in the ordinary Cartesian coordinates ( Fig.1 ).
The decrease comes from the expansion of the glasma.  

\begin{figure}[htb]
 \centering
 \includegraphics*[width=65mm]{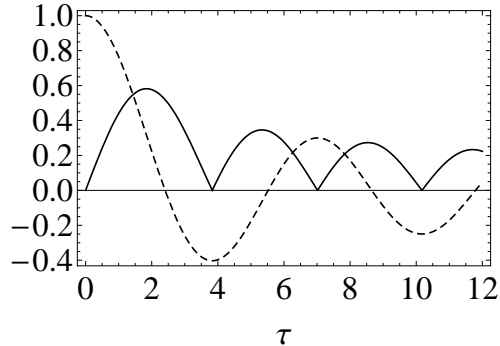}
 \caption{electric field $E(\tau)$ (dashed line)
and number density of electrons $N(\tau)$ (solid line) with arbitrary scale}
    \label{l2ea4-f1}
\end{figure}

The implication of the figure is in the following.
When the electric field $E_{\eta}>0$ is switched on, the pair production arises and 
the number density $N$ increases, while
the field becomes weak owing to the energy lose. 
The energy lose is caused from the acceleration of the charged particles. 
There are two effects which make the number density increase or decrease.
The number density increases by the pair production ( $s_{\eta} E_{\eta}$ ), 
while the expansion ( $-N/\tau $ ) makes the number density decrease:
$\partial_{\tau}N=s_{\eta} E_{\eta} -N/\tau$. 
At the beginning of the
pair production, $N$ increases 
because the effect of the pair production is stronger than
that of the expansion. But, at the time $\tau=\tau_0$,
$N$ stops increasing because both effects balance with each other.
After that, the effect of the expansion becomes stronger than
that of the pair production. Thus,
$N$ begins to decrease. On the other hand, the electric field 
loses its energy with the acceleration of the particles and vanishes 
at the time $\tau=\tau_1 > \tau_0$; $E_{\eta}(\tau=\tau_1)=0$.
Then, the field changes its direction, i.e. $E_{\eta}<0$.
The particles begin to be accelerated in the direction opposite to their velocity $k_{\eta}>0$.
Thus, the pair annihilation begins and makes the number density decrease furthermore.
Because the electric field accelerates the particles 
into the direction opposite to
the particle velocity, 
the field gains the energy from the particles.
Thus, the field becomes strong.
Eventually the number $N$ of the particles vanishes  ( $N(\tau=\tau_2)=0$ )
at the time $\tau=\tau_2>\tau_1$  when the annihilation stops
and the increase of the field strength also stops. We can see that
at the time $\tau=\tau_2$ a new pair production begins to take place under the effect 
of the electric field $E_{\eta}(\tau=\tau_2)<0$.
This is the physical explanation of the behaviors depicted in the figure 2.

We should also mention that the life time $\tau_1\simeq 2.4/\sqrt{2es_{\eta}}$ ( the first zero point of 
the field $E\propto J_0(\sqrt{2es_{\eta}}\tau)$ ) of the electric field is larger than
the corrensponding one $ t_1\equiv \pi/2\sqrt{2es}$ of $E\propto \cos(\sqrt{2es}\,t)$ in the non-expanding glasma
with $s=s_{\eta}$ or $B=B_{\eta}$.
Since the expansion makes the number density of the charged particles after their production lower
than that in the non-expanding case, the rate of the energy loss of the field 
is slower than that in the non-expanding case.
Hence, the life time $\tau_1$ is larger than $t_1$.

\section{summary and discussion}
\label{5}


To summarize, we have shown that due to the pair production of the charged particles,
the electric field homogeneous in $\eta$ decreases with the oscillation in
the $\tau$ and $\eta$ coordinates. 
This should be contrasted with the case in
the Cartesian coordinate where the field homogeneous in $z$ simply oscillates without 
the decrease. 
The number density of the electrons shows the similar behavior 
as that of the electric field. Since $\tau$ describes a proper time
associated with the expanding fluid, it is natural that
the energy of the electric field 
and the number density become lower with time $\tau$ for $\eta$ fixed.
This lower number density of the charged particles leads to the longer life time of
the electric field.
We expect that these behaviors shown in QED
also arise in the real glasma
produced in high energy heavy ion collisions.

We have only discuss the quark pair production
under the color electric field 
produced in high energy heavy ion collisions. 
In the collision the most important products would be gluons,
which can be also produced in the Schwinger mechanism\cite{itanji}.
In order to discuss the gluon production,
we need to take into account of the effect of 
the color magnetic field. 
As is well known, the gluons are unstable under the magnetic field.
It means that the gluons are coherently produced\cite{iwazaki4}.
Just as Higgs field $\phi_H$ located at the top of the potential
$\phi_H=0$, the gluon fields exponentially grow to approach the 
bottom of the potential.
The process can be regarded as the coherent production of the gluons.
It is important to find the ratio of the number density of the gluon to that of the quarks
produced in the Schwinger mechansm.
In near future, 
we discuss the gluon production as well as the quark production by
taking account of the back reaction of the coherent gluons and the quarks.

\vspace*{1cm}
We would like to
express thanks to Dr. K. Fukushima for making me pay attention to the recent paper
by Berges and Sexty.


\end{document}